\documentclass[aps,prb,twocolumn]{revtex4-1}
\usepackage[T1]{fontenc}
\usepackage[latin1]{inputenc}
\usepackage{lmodern}
\expandafter\let\csname equation*\endcsname\relax
\expandafter\let\csname endequation*\endcsname\relax
\usepackage{amsmath}
\usepackage{esint}
\usepackage{amssymb}
\usepackage{graphicx}
\usepackage{xcolor}
\usepackage{natbib}
\usepackage{bm}
\usepackage{bbold}
\usepackage[mathscr]{euscript}
\usepackage[cal=boondoxo]{mathalfa}
\usepackage{comment}

\def\ln{\textrm{ln}}

\def\pM{\mathrel{\raise 2pt \hbox{\tiny(}\!\raise 1pt \hbox{+}\settowidth {\dimen03} {+}\hskip-\dimen03 \raise -2.4pt \hbox {$-$} \!\raise 2pt \hbox{\tiny)}}}

\begin{document}
\title{Electric backaction on moir\'e mechanics}
\author{H\'ector Ochoa}
%\email{ho2273@columbia.edu}
\affiliation{Donostia International Physics Center, 20018 Donostia-San Sebastian, Spain\\
IKERBASQUE, Basque Foundation for Science, Maria Diaz de Haro 3, 48013 Bilbao, Spain\\
Department of Physics, Columbia University, New York, NY 10027, USA}

%\date{\today}
\begin{abstract}
A lattice mismatch between Van der Waals layers produces a moiré pattern and a subsequent electron band reconstruction. When the bilayer is charged, the sliding motion of one layer with respect to the other produces electric pumping. Here I discuss the reciprocal process: that a voltage bias produces a layer-shear mechanical force. The effect is deduced  from the lowest-order correction to the mechanical action by the coupling with electrons in an external field. In twisted bilayer graphene the new mechanical force is shown to be perpendicular to the applied field (due to C$_2$ symmetries exchanging the layers) and proportional to the charge density measured from neutrality. This is strictly true when the chemical potential is within a gap opened by the moir\'e potential due to a topological quantization, and approximately true when the chemical potential crosses the flat bands in a model including the self-consistent Hartree interaction. %This is the result of the charge density adiabatically following smooth (on the scale of the moiré period) deformations of the stacking order.
%When the relative position of the layers is locked a long-wavelength deformation of the moir\'e pattern contributes to partially screen in-plane electric fields. The reciprocal effect
In a mechanical device the effect should be manifested as an apparent enhancement of the friction between layers when the system is charged. %when they slide but charge in the moir\'e pattern cannot flow.
Depinning fields for the sliding motion of the layers are estimated in the order of $\mathcal{E}_c\sim$ kV/cm around the magic angle. %Altogether the conclusion is that the electric control of the stacking order in nonpolar moir\'e superlattices is viable although experimentally challenging.
\end{abstract}
\maketitle

\section{Introduction}

The moir\'e patterns produced by a small lattice mismatch between Van der Waals layers reorganize their electronic spectrum in narrow bands defined on a smaller Brillouin zone. In twisted bilayer graphene\cite{portu} and other carbon-based structures with an emergent 6-fold symmetry\cite{magic} this process is optimal at certain magic angles.\cite{Morell_etal,BM} At low temperatures the flat bands can host broken-symmetry phases and physics of correlated electrons yet to be understood.\cite{Cao1,Cao2,Columbia,Sharpe_etal,Efetov,stm1,stm2,stm3,stm4,cascade1,Serlin_etal,cascade1,cascade2,Cao3,pom1,pom2} The quantum geometry of the flat bands is likely to play a fundamental role, in particular for the stability of the superconducting phase.\cite{sc1,sc2,sc3}

Another manifestation of geometrical phases of electrons in moir\'e patterns is the prediction of a charge pumping as the result of the translation of one layer with respect to the other \cite{pumping1,pumping2,pumping3} (see also Ref.~\onlinecite{edge3}). For a bilayer with a 3-fold or 6-fold principal axis (along $\hat{\mathbf{z}}$), the pumped electric current density is related to the sliding velocity $\boldsymbol{v}$ by %\cite{pumping3}
\begin{align}
\label{eq:constitutive1}
\boldsymbol{j}=C_{\parallel}\,\boldsymbol{v}+C_{\perp}\,\hat{\mathbf{z}}\times\boldsymbol{v},
\end{align}
where $C_{\parallel,\perp}$ are phenomenological parameters constrained by symmetry. For example, in the case of homobilayers, where the moir\'e pattern results from a twist, additional 2-fold rotation axes within the plane of the device impose $C_{\parallel}=0$. In the case of heterobilayers with no relative twist, where the moiré pattern results from the difference in lattice parameters, additional mirror reflection planes containing $\hat{\mathbf{z}}$ yield $C_{\perp}=0$.

As pointed out in Ref.~\onlinecite{pumping3}, Eq.~\eqref{eq:constitutive1} is a general phenomenological relation that should be valid in any situation. Nevertheless, the topological character of this response is only manifest when the chemical potential lies within a gap opened by the moir\'e potential. The authors of Refs.~\onlinecite{pumping1,pumping2,pumping3} show independently that when the sliding motion is slow enough compared to the time scale set by the inverse of the electronic gap twisted bilayer graphene realizes a two-dimensional analogue of a Thouless pump.\cite{Thouless} 

Here I provide a complementary discussion of this effect by focusing on the reciprocal process. I show that a voltage bias makes one layer to slide with respect to the other; more specifically, in the presence of an electromotive force $\boldsymbol{\mathcal{E}}$ there is a mechanical force (density) between layers of the form\begin{align}
\label{eq:constitutive2}
\boldsymbol{f}=C_{\parallel}\,\boldsymbol{\mathcal{E}}-C_{\perp}\,\hat{\mathbf{z}}\times\boldsymbol{\mathcal{E}}.
\end{align}
This relation is reciprocal to the pumping current in Eq.~\eqref{eq:constitutive1} and is therefore governed by the same phenomenological constants.

The structure of the manuscript is as follows. First I will compute the correction to the mechanical action describing the dynamics of stacking configurations in the presence of an external field; this is Sec.~\ref{sec:force} with some technical details relegated to appendices. In the limit of weak fields and smooth stacking fluctuations the phenomenological parameters describing this coupling are given by integrals of the sliding Berry curvature introduced in Refs.~\onlinecite{pumping1,pumping2,pumping3}. These are evaluated for the continuum model of twisted bilayer graphene at the magic angle including also the effect of a self-consistent Hartree potential. The force is shown to be proportional to the charge density measured from neutrality, although this result is only rigorous when the chemical potential lies within a mini-band gap (i.e., only when the adiabatic regime can be rigorously defined). I will write general constitutive relations for the coupled charge and stacking dynamics and I will apply them to different experimental scenarios in Sec.~\ref{sec:constitutive}. This analysis shows that if the relative position of the layer is locked (for example, due to the presence of the contacts) there is still a negative correction to the electric polarization related to the mechanical deformation of the moir\'e pattern due to pinning forces that contribute to partially screen the external field when the chemical potential is within a gap. The reciprocal effect is an effective increase in the friction between layers when the moir\'e pattern slides but charge cannot flow. Finally, I will show in Sec.~\ref{sec:depinning} that for weak disorder, the depinning fields for the sliding motion of the layers are estimated in $\mathcal{E}_c\approx 5$ kV/cm for magic-angle graphene. These are larger, for example, than the typical threshold fields for sliding conduction of charge density waves.\cite{RMP} I will conclude in Sec.~\ref{sec:conclusions} with a discussion on the experimental viability of controlling the stacking order by the application of electric fields.

\section{Adiabatic force on stacking order}

\label{sec:force}

\subsection{Preliminaries: Space of stacking configurations}

\begin{figure}
\centerline{\includegraphics[width=\columnwidth]{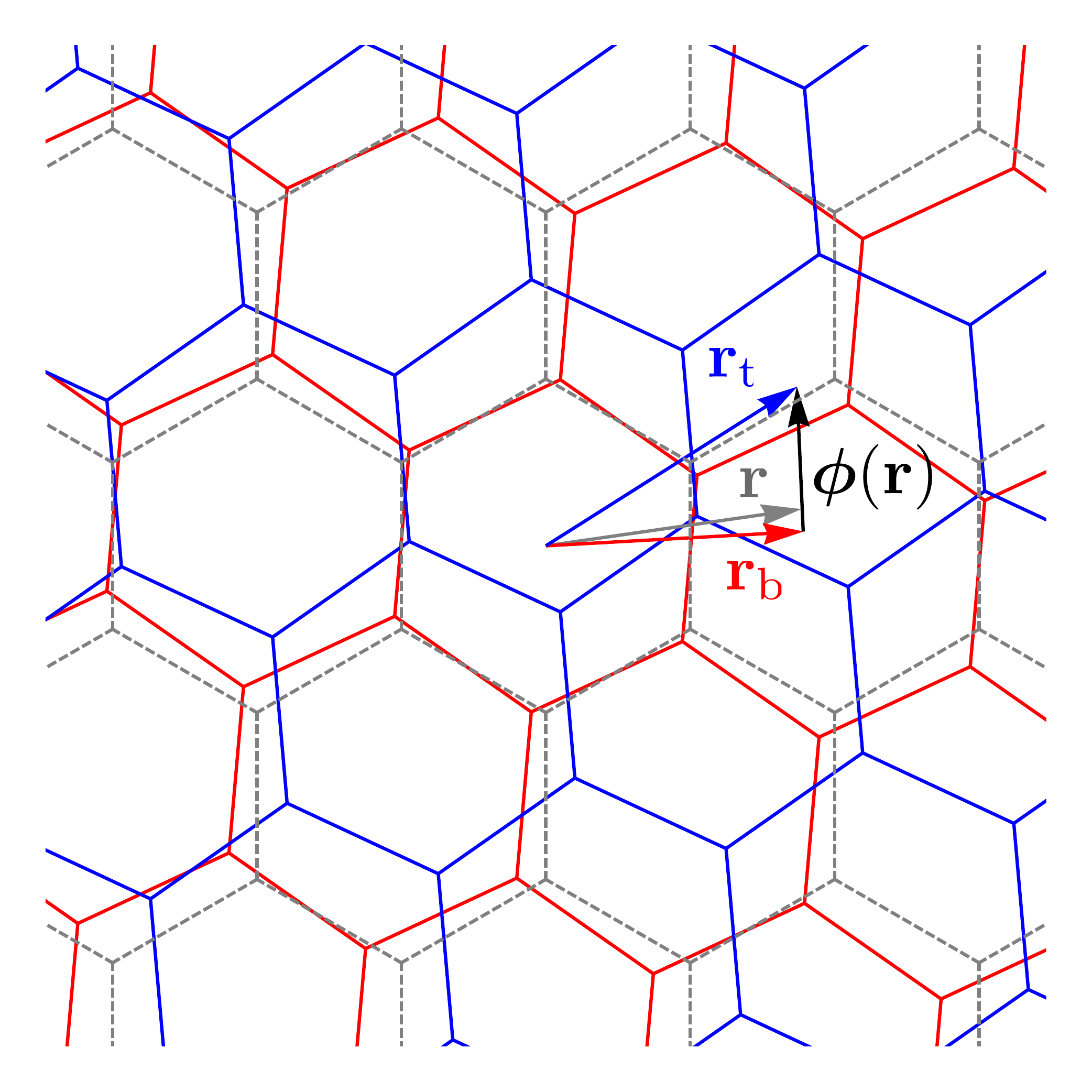}}
\caption{\textbf{Parametrization of stacking configurations}. The field $\boldsymbol{\phi}(\mathbf{r})$ relates a position $\mathbf{r}$ in space with the local stacking configuration given as the relative translation of the layers that would generate the same configuration starting from maximum lattice overlap (AA stacking in graphene bilayers). Consider then a position $\mathbf{r}$ (in gray) and its projections in the top (blue) and bottom (red) layers before the shift, in this case a rigid rotation followed by a rigid translation of the layers; $\boldsymbol{\phi}(\mathbf{r})$ is defined as the difference between the projections after the shift, $\boldsymbol{\phi}(\mathbf{r})=\mathbf{r}_{\textrm{t}}-\mathbf{r}_{\textrm{b}}$.}
\label{fig:fig1}
\end{figure}

The first assumption in the following derivation is that the electronic Hamiltonian of the bilayer in the absence of external fields only depends on positions through its parametric dependence on the local stacking configuration, $\hat{\mathcal{H}}[\boldsymbol{\phi}(\mathbf{r})]$, represented by the vector-valued field $\boldsymbol{\phi}(\mathbf{r})$; see Fig.~\ref{fig:fig1} for a definition. The sliding velocity introduced before measures the rate of changes in stacking configurations, $\boldsymbol{v}=\delta\dot{\boldsymbol{\phi}}$. The important observation is that this space is periodic by construction: vectors $\boldsymbol{\phi}$ and $\boldsymbol{\phi}+\mathbf{R}$, where $\mathbf{R}$ is a Bravais vector of graphene's lattice, represent the same stacking configuration.

The notion of a moir\'e period follows from this observation: At small twist angles the local stacking configuration corresponding to a rigid rotation of angle $\theta$ can be approximated by $\boldsymbol{\phi}_0(\mathbf{r})\approx \theta\, \hat{\mathbf{z}}\times\mathbf{r}$,\cite{foot_rot} and hence $\boldsymbol{\phi}_0(\mathbf{r}+\mathbf{R}_{\textrm{m}})=\boldsymbol{\phi}_0+\mathbf{R}\equiv\boldsymbol{\phi}$ by construction, where $\mathbf{R}_{\textrm{m}}=\theta^{-1}\,\mathbf{R}\times\hat{\mathbf{z}}$ are the vectors of the moir\'e pattern in real space. More generally, the equilibrium stacking texture $\boldsymbol{\phi}_0(\mathbf{r})$ (including lattice relaxation) describes a mapping between real space and the manifold of stacking configurations, \begin{align}
\label{eq:mapping}
\boldsymbol{\phi}_0\left(\mathbf{r}\right):\mathbf{r}\longrightarrow \boldsymbol{\phi}\in S_1\times S_1\equiv T_2,
\end{align}
which possesses the topology of a torus. The simplest mechanism for the mechanical force in Eq.~\eqref{eq:constitutive2} is provided by the accumulation of geometrical phases by the electronic wave function along a non-trivial loop defined on this space.

\subsection{Effective action}

The dynamics of the stacking field $\boldsymbol{\phi}(t,\mathbf{r})$ is governed by the Lagrangian\begin{align}
\mathcal{L}_{\textrm{mech}}\left[\boldsymbol{\phi}(t,\mathbf{r})\right]=\frac{\varrho_{\phi}}{2}\int d\mathbf{r}\,\dot{\boldsymbol{\phi}}^2-\mathcal{U}\left[\boldsymbol{\phi}(t,\mathbf{r})\right],
\end{align}
where $\mathcal{U}[\boldsymbol{\phi}(t,\mathbf{r})]$ is the mechanical free energy of the bilayer and $\varrho_{\phi}$ parametrizes the inertia of the relative motion of the two layers, hence $\varrho_{\phi}=\varrho/2$, where $\varrho$ is the areal mass density of each layer.

The form of $\mathcal{U}$ is not relevant for the following discussion but it consists essentially of two competing terms, one accounting for the adhesion forces between layers favoring certain stacking configurations, and the other penalizing intralayer stress. The equilibrium stacking texture $\boldsymbol{\phi}_0(\mathbf{r})$ is the result of the minimization of $\mathcal{U}$ subjected to a boundary condition that stabilizes the twist angle $\theta$. Putting aside the important question of the stability of the angle as the layers slide, I will consider then the dynamics around this equilibrium solution, $\boldsymbol{\phi}(t,\mathbf{r})=\boldsymbol{\phi}_0(\mathbf{r})+\delta \boldsymbol{\phi}(t,\mathbf{r})$. In particular, we are interested in the slow sliding dynamics of the bilayer assuming that the layers are free to slide, i.e., that the layers are incommensurate to each other so that the sliding motion adiabatically connects isoenergetic stacking configurations involving no distortion of the moir\'e pattern. The local change in stackings is of the form $\delta\boldsymbol{\phi}(t,\mathbf{r})=\boldsymbol{\phi}_0(\mathbf{r}-\boldsymbol{u}(t))-\boldsymbol{\phi}_0(\mathbf{r})$, where $\boldsymbol{u}(t)$ parametrizing the adiabatic path in stacking configurations corresponds to the collective coordinate associated with soft collective modes (phasons) in the spectrum of stacking fluctuations.\cite{Koshino,phasonsI,phasons_TMD,phasons,phasons_Eslam,phasons_Matthias} If we neglect lattice relaxation this is just a rigid translation of one layer with respect to the other. More generically, it parametrizes the travelling wave associated with the sliding motion of stacking domain walls resulting from the relaxation process. %The dependence of the electronic spectrum (and of the Berry curvatures below) in $\boldsymbol{u}$ is absent provided that it can be gauged away from the Hamiltonian by a shift in coordinates.

The goal here is to compute the correction to the mechanical action in the presence of external fields. We start with the action for electrons in the bilayer coupled to a scalar ($V$) and vector ($\mathbf{A}$) potentials,\begin{align}
S\left[\psi,\psi^{\dagger}\right]=\int dt\int d\mathbf{r}\,\psi^{\dagger}\left(i\hbar\partial_t+eV+\mu-\hat{\mathcal{H}}\left[\boldsymbol{\phi},\mathbf{A}\right]\right)\psi,
\end{align}
where $\mu$ is the chemical potential. The vector potential enters in minimal coupling, $-i\hbar\boldsymbol{\partial}\rightarrow -i\hbar\boldsymbol{\partial}+e\mathbf{A}/c$. We can formally integrate out electrons, $iS_{eff}[\boldsymbol{\phi}]/\hbar=\textrm{Tr}\,\ln\hat{\mathcal{G}}[\boldsymbol{\phi}]$, to produce an effective action for stacking configurations through the parametric dependence of the Hamiltonian; here $\hat{\mathcal{G}}[\boldsymbol{\phi}]=(i\hbar\partial_t+eV+\mu-\hat{\mathcal{H}}\left[\boldsymbol{\phi},\mathbf{A}\right])^{-1}$ and the trace represents the summation over all quantum numbers.

Since we are interested only in the linear response to external fields and the stacking configurations are assumed to be smoothly varying we can expand the logarithm in first order in the fields and in first derivatives of the stacking fluctuations $\delta\boldsymbol{\phi}$. Diagrammatically, this second-order contribution to the effective action corresponds to the usual polarization bubble with the electron-phason coupling $\partial_{\phi}\hat{\mathcal{H}}$ in one of the vertices. After expanding in linear order to the external momenta/frequency, the final result can be recast as\begin{align}
S_{eff}^{(2)}\left[\delta\boldsymbol{\phi}\right]=\tilde{C}_{ij}\int dt\int d\mathbf{r}\, \left(eV\partial_i\delta\phi_j+\frac{e}{c}A_i\delta\dot{\phi}_j\right),
\end{align}
where repeated indices are summed up, and the coefficients $\tilde{C}_{ij}$ have the form of a Pontryagin index,\begin{align}
\label{eq:index}
\tilde{C}_{ij}=\int\frac{dq}{(2\pi)^3}\,\textrm{Tr}\left[\hat{\mathcal{G}}_0\cdot\partial_{\omega}\hat{\mathcal{G}}_0^{-1}\cdot\hat{\mathcal{G}}_0\cdot\partial_{\phi_j}\hat{\mathcal{G}}_0^{-1}\cdot \hat{\mathcal{G}}_0\cdot\partial_{q_i}\hat{\mathcal{G}}_0^{-1}\right],
\end{align}
with $q=(\omega,\mathbf{q})$. Here I have introduced the \textit{free} Green operator in the absence of external fields and with the Hamiltonian evaluated at the equilibrium stacking configuration, $\hat{\mathcal{H}}_0\equiv\hat{\mathcal{H}}[\boldsymbol{\phi}_0])$; thus\begin{align}
\hat{\mathcal{G}}_0\left(\omega,\mathbf{q}\right)=\left[\hbar\omega+i0^+\textrm{sign}(\omega)+\mu-\hat{\mathcal{H}}_0\left(\mathbf{q}\right)\right]^{-1}.
\end{align}
In the last expression I have introduced the Fourier transform of the electronic Hamiltonian with momenta $\mathbf{q}$ defined within the moir\'e Brillouin zone (mBZ).

$S_{eff}^{(2)}$ is the lowest-order correction in derivatives of the stacking field to the mechanical action. From this term we see that the field produces a force (density) of the form\begin{align}
\boldsymbol{f}=\frac{\delta S_{eff}^{(2)}}{\delta\boldsymbol{\phi}}=\hat{C}^T\cdot\boldsymbol{\mathcal{E}},
\end{align} 
where $\boldsymbol{\mathcal{E}}=-\boldsymbol{\nabla}V-\dot{\mathbf{A}}/c$ is the electric field and $\hat{C}$ is a matrix with elements $C_{ij}=e\,\tilde{C}_{ij}$. This is nothing but Eq.~\eqref{eq:constitutive2}. The symmetry constraints discussed in the introduction follows from the symmetry transformations of momentum and stacking fields. Both are vectors but note, in particular, that they have opposite signature with respect to operations exchanging the layers.

By the same token, there is a contribution to the electric current (density) given by\begin{align}
\boldsymbol{j}=c\frac{\delta S_{eff}^{(2)}}{\delta\mathbf{A}}=\hat{C}\cdot\delta\dot{\boldsymbol{\phi}},
\end{align}
which is another way to write Eq.~\eqref{eq:constitutive1}. The constants $C_{ij}$ are the same in both expressions but the corresponding matrices are transposed. %This is in fact an example of Onsager reciprocity relations.

Finally, performing the integral in frequencies in Eq.~\eqref{eq:index} (see Appendix~\ref{sec:app1}), the response coefficients can be written as\begin{align}
\label{eq:sliding_Berry}
C_{ij}= e\sum_{\alpha,n}\int_{\textrm{mBZ}}\frac{d\mathbf{q}}{(2\pi)^2}\,\Omega_{q_i\phi_j}^{(n,\alpha)}\left(\mathbf{q}\right)\,\Theta\left(\mu-\varepsilon_{n,\alpha}(\mathbf{q})\right),
\end{align}
where $\Omega_{q_i\phi_j}^{(n,\alpha)}(\mathbf{q})$ the sliding Berry curvature introduced in Ref.~\onlinecite{pumping1,pumping2,pumping3} summed over all occupied states with energy dispersion $\varepsilon_{n,\alpha}(\mathbf{q})$.

\subsection{Application to twisted bilayer graphene}

\subsubsection{Non-interacting bands}

\begin{figure}
\centerline{\includegraphics[width=\columnwidth]{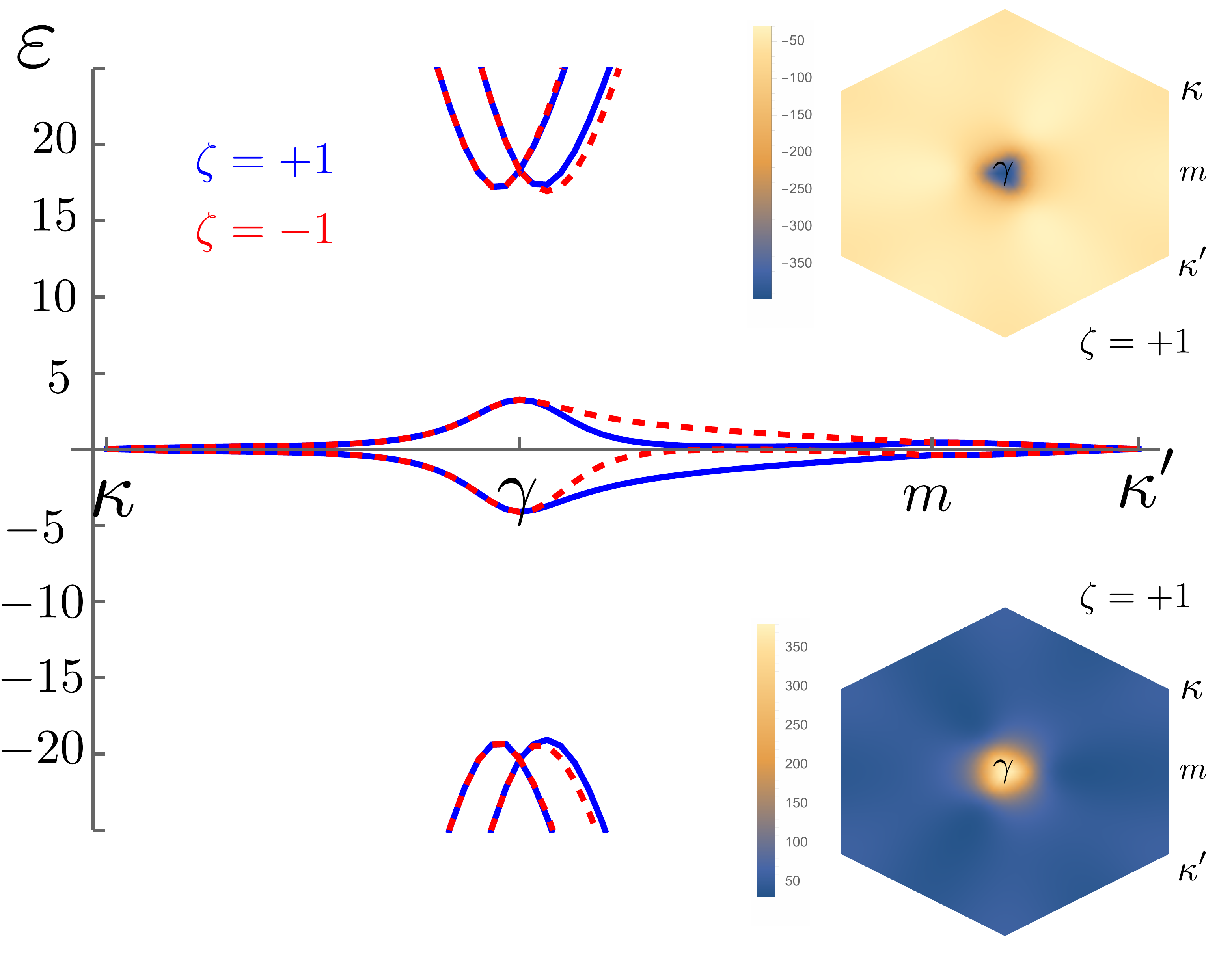}}
\caption{\textbf{Flat bands and sliding Berry curvatures at the magic angle}. The insets show the distribution in momentum space of $\Omega_{\perp}^{(n,+)}(\mathbf{q})$ in Eq.~\eqref{eq:symmetric_Berry} summed over occupied bands when the highest flat band is fully occupied (top) and when the lowest flat band is fully empty (bottom). I truncated the continuum Hamiltonian to 91 plane waves for each layer and sublattice.}
\label{fig:fig2}
\end{figure}

Next I evaluate the matrix of coefficients $C_{ij}$ within the continuum model of twisted bilayer graphene.\cite{portu,BM} The model consists of two Dirac Hamiltonians describing the original linear dispersion of the bands on each layer at valleys centered at the two inequivalent points $\mathbf{K}_{\pm}=\pm\mathbf{K}_+$ of the hexagonal Brillouin zone. Tunneling from the bottom to the top layer is described by the term\begin{align}
\hat{T}_{\zeta}\left(\mathbf{r}\right)=\sum_{n=0}^2 e^{i\zeta\left(\mathbf{q}_n+\mathbf{K}_+\right)\cdot\boldsymbol{\phi}\left(\mathbf{r}\right)}\, e^{i\zeta\frac{2\pi}{3}\hat{\sigma}_z} \hat{T}_{0}\,e^{-i\zeta\frac{2\pi}{3}\hat{\sigma}_z},
\end{align}
where $\zeta=\pm 1$ labels valleys $\mathbf{K}_{\pm}$, and $\mathbf{q}_{0,1,2}=\mathbf{0},\mathbf{g}_2,-\mathbf{g}_1$, where $\mathbf{g}_{1,2}$ are primitive vectors of graphene's reciprocal lattice forming 120$^{\textrm{o}}$. Operators $\hat{\sigma}_i$ are Pauli matrices acting on the sublattice degree of freedom of the spinor wave functions. The matrix $\hat{T}_0$ contains the tunnellng rates; a parametrization compatible with $D_6$ symmetry is of the form\begin{align}
\hat{T}_0=t_{\textrm{AA}}\,\hat{1}+t_{\textrm{AB}}\,\hat{\sigma}_x.
\end{align}
Plugging $\boldsymbol{\phi}_0(\mathbf{r})\approx \theta\, \hat{\mathbf{z}}\times\mathbf{r}$ into these equations one recovers the usual expression of the continuum model with moir\'e reciprocal vectors given by $\mathbf{G}_i=\theta^{-1}\mathbf{g}_i\times\hat{\mathbf{z}}$. Lattice relaxation produces more harmonics but I will neglect those and take slightly different values of the interlayer tunneling rates at different stacking regions, $t_{\textrm{AA}}=79.7$ meV and $t_{\textrm{AB}}=97.5$ meV.\cite{Koshino2} Figure~\ref{fig:fig2} shows the lowest-energy bands at the magic angle ($\theta=1.05^{\textrm{o}}$ for these model parameters) including up to 91 plane waves in the calculation for each layer and sublattice.

As discussed in the introduction, $D_6$ symmetry imposes $C_{xx}=C_{yy}=0$ and $C_{yx}=-C_{xy}\equiv C_{\perp}$, the latter given by the integral of the corresponding component of the sliding Berry curvature. The insets in Fig.~\ref{fig:fig2} show the distribution of the symmetric combination \begin{align}
\label{eq:symmetric_Berry}
\Omega_{\perp}^{(n,\zeta)}\left(\mathbf{q}\right)=\frac{1}{2}\Omega_{q_y\phi_x}^{(n,\zeta)}\left(\mathbf{q}\right)-\frac{1}{2}\Omega_{q_x\phi_y}^{(n,\zeta)}\left(\mathbf{q}\right)
\end{align}
for valley $\zeta=+1$, summed over the 182 bands below the flat bands (lower inset), and including also the flat bands (upper inset). The distribution in the opposite valley follows from time-reversal symmetry, $\Omega_{\perp}^{(n,-)}\left(\mathbf{q}\right)=\Omega_{\perp}^{(n,+)}\left(-\mathbf{q}\right)$. Summing $\Omega_{\perp}^{(n,+)}(\mathbf{q})$ on a grid of $1933$ $\mathbf{q}$-points and accounting for the degeneracy 4 from spin and valley I obtained $A_{\textrm{m}}C_{\perp}/4e=56.31$ for the lower inset and $A_{\textrm{m}}C_{\perp}/4e=-56.52$ for the upper inset ($A_{\textrm{m}}=\sqrt{3}a^2/2\theta^2$ is the area of the moir\'e unit cell in real space, where $a$ is graphene's lattice constant). These are numerically very close to $\pm \theta^{-1}=\pm 54.57$. %As the twist angle increases (the rest of parameters of the calculation being the same) the numerical accuracy to $\pm \theta^{-1}$ improves.
Indeed, I checked that $C_{\perp}=\pm 4e/\theta A_{\textrm{m}}$ for different values of twist angle within the same numerical error (results not shown), or simply in terms of the electron density $n$ measured form neutrality:

\begin{align}
\label{eq:estimate}
C_{\perp}=-\theta^{-1}en=-\frac{8e\,\theta}{\sqrt{3}a^2}.
\end{align}
The same estimate (with opposite sign) follows for hole dopings.

\subsubsection{Hartree bands}

As it is apparent from the insets in Fig.~\ref{fig:fig2} most of the sliding Berry curvature of the flat bands is concentrated around the the zone center ($\gamma$ point in the figures), so $C_{\perp}$ does not necessarily track the carrier density as the flat bands are gradually populated. However, it is well known that electrostatic effects alter significantly the dispersion of the flat bands as charge is added or removed from the bilayer.\cite{Hartree1,Hartree2,Hartree3,Hartree4} In order to investigate charging effects in the bands and the distribution of sliding Berry curvature, I included in the electronic Hamiltonian the Coulomb repulsion among electrons treated in a self-consistent Hartree approximation. Details can be found in Appendix~\ref{sec:app2}.

\begin{figure}
\centerline{\includegraphics[width=\columnwidth]{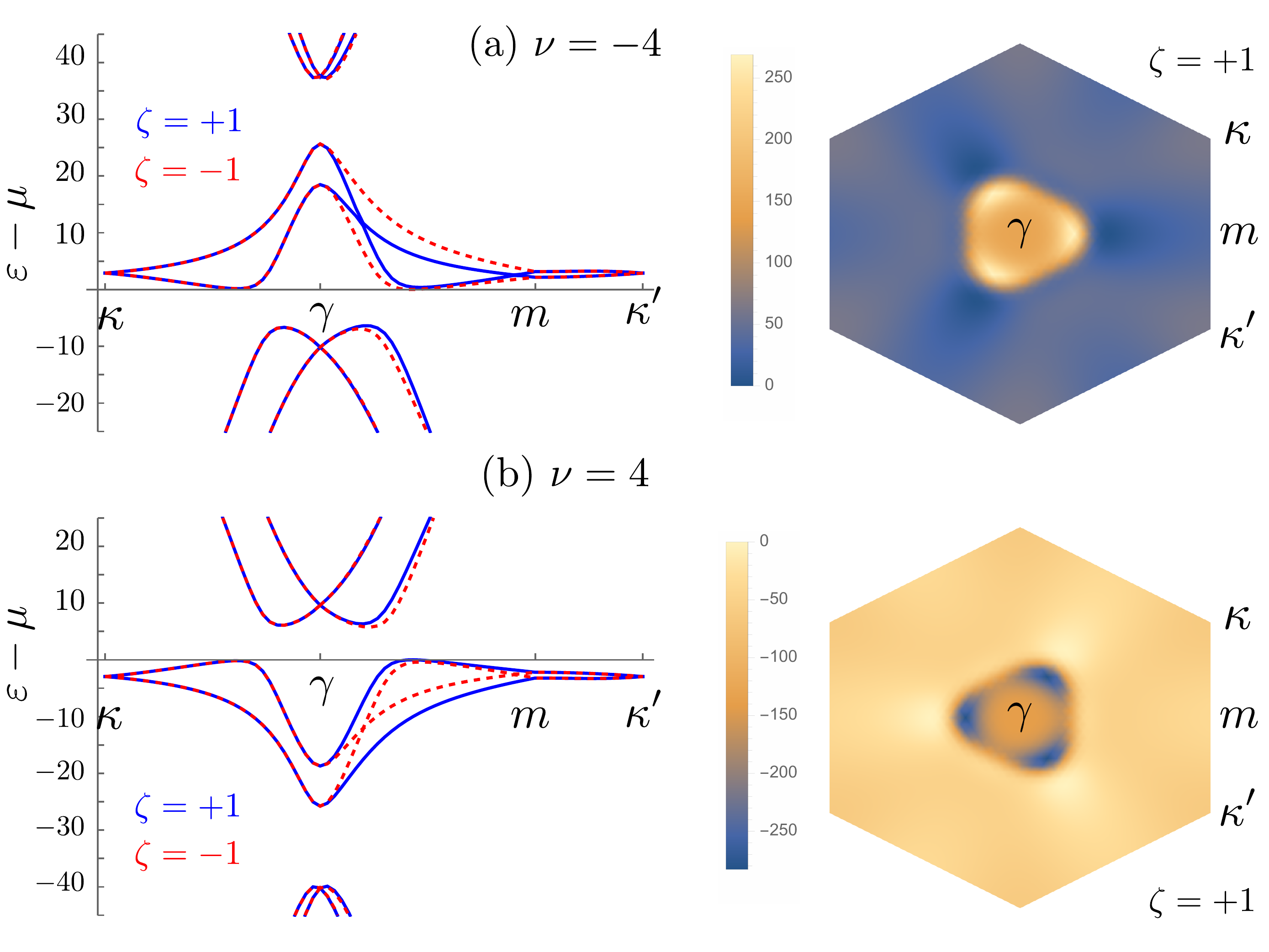}}
\caption{\textbf{Hartree bands and sliding Berry curvature}. The dielectric constant is $\epsilon_r=7.5$. I took the same number of plane waves as in the calculations in Fig.~\ref{fig:fig2}, and the harmonics of the Hartree potential were restricted to the first star. (a) Filling $\nu=-4$, when the chemical potential lies within the gap between the lowest flat band and the rest of the spectrum. The right panel shows $\Omega_{\perp}^{(n,+)}(\mathbf{q})$ summed over all occupied bands. (b) Filling $\nu=4$, when the chemical potential lies within the gap between the highest flat band and the rest of the spectrum. The right panel shows $\Omega_{\perp}^{(n,+)}(\mathbf{q})$ summed over all occupied bands.}
\label{fig:fig3}
\end{figure}

Figure~\ref{fig:fig3} shows the self-consistent Hartree band structures corresponding to a long-range Coulomb interaction with a relative permittivity $\epsilon_r=7.5$ for fully empty (panel a) and fully occupied (panel b) flat bands. The density plots represent $\Omega_{\perp}^{(n,+)}\left(\mathbf{q}\right)$ summed over all filled bands for each filling. The sum in the same momentum grid as before gives $A_{\textrm{m}}C_{\perp}/4e=55.66$ for filling $\nu=-4$ and $A_{\textrm{m}}C_{\perp}/4e=-55.69$ for filling $\nu=4$. 
\subsubsection{$C_{\perp}$ as a function of band filling}

\begin{figure}
\centerline{\includegraphics[width=\columnwidth]{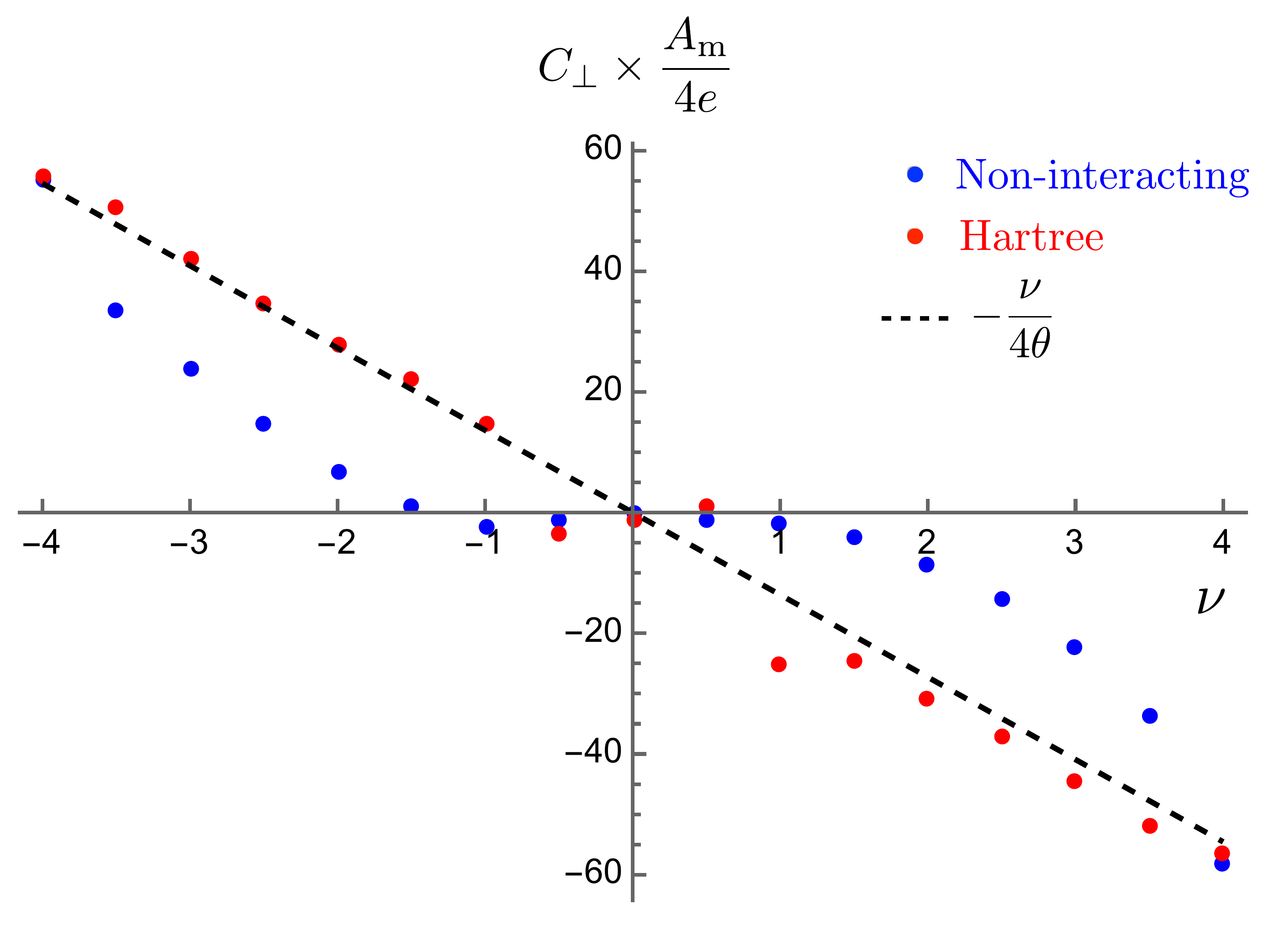}}
\caption{\textbf{$C_{\perp}$ as a function of filling of the flat bands}. Blue dots correspond to the non-interacting theory, red dots to the self-consistent Hartree theory. The dashed black line corresponds to $C_{\perp}$ tracking the charge density.}
\label{fig:fig4}
\end{figure}

The calculations in Fig.~\ref{fig:fig3} show that the sliding Berry curvature with the account of the Hartree potential is in fact more spread over the moir\'e Brillouin zone. How do these changes affect the value of $C_{\perp}$ as the Fermi level crosses the flat bands? Here I am concerned only about the magnitude of the reactive mechanical force induced by the field, although we are not in the adiabatic regime and we should expect also dissipative terns due to electron quasiparticle excitations.

The answer is in Fig.~\ref{fig:fig4}, which shows the numerical evaluation of $C_{\perp}$ as a function of filling $\nu$ running from $-4$ (completely empty flat bands) to $+4$ (completely full). The blue dots correspond to the rigid non-interacting bands, in which $C_{\perp}$ decreases rapidly as the lowest flat band starts to be populated, then it is roughly $0$ for a wide range of densities around the neutrality point, and then decreases quickly again.

However, for the non-rigid Hartree bands, which change with filling, the behavior is different; these correspond to the red points in Fig.~\ref{fig:fig4}. We can see that $C_{\perp}$ is roughly proportional to charge density (dashed black line for reference) except for a much narrower window of fillings around neutrality. This is the result of the filling-dependent change in band dispersion associated with the smoothening of the charge density in real space together with the spreading of the sliding Berry curvature in reciprocal space.

\subsubsection{Coupled charge and twist-angle fluctuations}

The quantization of $C_{\perp}$ in units of $e/\theta A_{\textrm{m}}$ when the Fermi level lies within a gap can be understood by rewritting Eq.~\eqref{eq:sliding_Berry} in terms of the four Chern numbers defined on manifolds $S_1\times S_1$ spanned by a primitive vector of the moir\'e Brillouin zone and a primitive vector of the stacking configuration space.\cite{pumping1,pumping2,pumping3} Physically, the integral of the sliding Berry curvature describes how the Wannier center of the band changes with stacking, so $C_{\perp}$ tracks the charge density in the adiabatic limit. 

Another way to understand the relation between $C_{\perp}$ and band filling is the following relation for the charge density fluctuation associated with spatial variations of the stacking configuration. From the effective action we have\begin{align}
\label{eq:charge_fluctuation}
\delta n=-\frac{\delta S_{\textrm{eff}}^{(2)}}{\delta V}=-C_{ij}\partial_i\delta\phi_j=C_{\perp}\left(\boldsymbol{\nabla}\times\delta\boldsymbol{\phi}\right)_{z}.
\end{align}
This relation is complementary to the expression of the pumping current, in the sense that the gauge invariance of the new term in the effective action guarantees that the pumped current is conserved, $\delta\dot{n}+\boldsymbol{\nabla}\cdot\boldsymbol{j}=0$.

If we compare the right-hand side with the stacking texture for a rigid rotation, $\boldsymbol{\phi}_0(\mathbf{r})\approx\theta\,\hat{\mathbf{z}}\times\mathbf{r}$, then we can read Eq.~\eqref{eq:charge_fluctuation} as a relation between adiabatic changes in the charge distribution and smooth (on the scale of the moir\'e pattern) spatial variations of the twist angle: $(\boldsymbol{\nabla}\times\delta\boldsymbol{\phi})_{z}\approx 2\delta\theta$, hence $\delta n=2 C_{\perp}\delta{\theta}$. On the other hand, for a fixed filling we can write $n=-e \nu/A_{\textrm{m}}\propto\theta^2$. A smooth change in twist angle modifies the area of the moir\'e cell and thus the density, $\delta n=\delta{\theta}\times\delta_{\theta} n=2n\delta{\theta}/\theta$. Comparing both expressions we expect then $C_{\perp}=n/\theta=-e\nu/\theta A_{\textrm{m}}$, which agrees with Eq.~\eqref{eq:estimate} for full fillings, $\nu=\pm 4$.

\section{Constitutive relations}

\label{sec:constitutive}

The purpose of this section is twofold: First I will show that regardless of its microscopic origin, the force in Eq.~\eqref{eq:constitutive2} follows from the phenomenological expression for the pumping current in Eq.~\eqref{eq:constitutive1} as a result of Onsager reciprocity. Then I will apply the constitutive relations to different experimental scenarios.

\subsection{Mechanical force from Onsager reciprocity}

The procedure is to write general constitutive relations describing the coupled charge and sliding dynamics of the bilayer. These expressions relate flows with their conjugate thermodynamic forces via a matrix of linear response functions. The product of the flows with their conjugated forces determine the work that we have to supply in order to sustain transport. When charge flows and one of the layers slides the work dissipated per unit of area is \begin{align}
\nonumber
\Delta \bar{W}=\int_{-\infty}^{\infty} \frac{d\omega}{2\pi}\, \left[\boldsymbol{j}^{*}\left(\omega\right)\cdot\boldsymbol{\mathcal{E}}\left(\omega\right)+\boldsymbol{v}^{*}\left(\omega\right)\cdot\boldsymbol{f}\left(\omega\right)\right].
\end{align}
Here I have introduced Fourier transforms in time of currents and forces. From this expression we identify the electric field $\boldsymbol{\mathcal{E}}$ as the thermodynamic force conjugate to the electric current. The force conjugate to the sliding velocity is simply $\boldsymbol{f}=-\delta\mathcal{U}/\delta\boldsymbol{\phi}$. In linear response we write\begin{align}
\left[\begin{array}{c}\boldsymbol{j}\left(\omega\right)\\
\boldsymbol{v}\left(\omega\right)
\end{array}\right]=\left[\begin{array}{cc}
\hat{\sigma}\left(\omega\right) & \hat{A}\left(\omega\right)\\
\hat{B}\left(\omega\right) & -i\omega\,\hat{\chi}\left(\omega\right)
\end{array}\right]\left[\begin{array}{c}\boldsymbol{\mathcal{E}}\left(\omega\right)\\
\boldsymbol{f}\left(\omega\right)
\end{array}\right].
\end{align}
%Currents and forces are related by a block matrix of dynamical response functions.
Thermodynamic stability requires that the real part of this block matrix is positive semi-definite. Time-reversal symmetry dictates that it must be symmetric; in particular\begin{align}
\label{eq:Onsager}
\hat{B}\left(\omega\right)=\hat{A}^{T}\left(\omega\right).
\end{align}
In the $\omega=0$ limit this property reduces to the Onsager reciprocal relations among transport coefficients.

Before jumping into the off-diagonal blocks containing $C_{\parallel,\perp}$, it is worth discussing the diagonal blocks for a moment. In the charge sector $\boldsymbol{j}(\omega)$ and $\boldsymbol{\mathcal{E}}(\omega)$ are related by the optical conductivity, $\hat{\sigma}(\omega)$. In the mechanical sector sliding velocities and forces are related by the long-wavelength ($\mathbf{q}=0$) limit of the stacking susceptibility tensor $\hat{\chi}\left(\omega,\mathbf{q}\right)$ introduced in Ref.~\onlinecite{phasons}. For the purposes of DC transport we only need to capture the behavior at small frequencies, which is dominated by the phason modes\cite{Koshino,phasonsI,phasons_TMD,phasons,phasons_Eslam,phasons_Matthias} introduced before. We can write then\cite{phasons}
\begin{align}
\label{eq:susceptibility}
\hat{\chi}\left(\omega,\mathbf{q}\right)=\sum_{\nu=L,T}\frac{\varrho_{\phi}^{-1}\hat{P}_{\nu,\mathbf{q}}}{\omega_{\nu,\mathbf{q}}^2-\omega^2-i\omega\gamma_{\nu}\left(\omega,\mathbf{q}\right)},
\end{align}
where $\hat{P}_{\nu,\mathbf{q}}$ are projectors on the space of phason modes labelled by momentum $\mathbf{q}$ within the moir\'e Brillouin zone (in a continuum elastic theory that assumes moir\'e translational invariance) and branch indices $\nu=L,T$. These modes correspond to longitudinal ($L$) and transverse ($T$) traveling waves of the moir\'e pattern. They are soft in the sense that $\omega_{\nu,\mathbf{q}}\rightarrow 0$ as $\mathbf{q}\rightarrow 0$ due to incommensurability (or negligible energy differences between commensurate and incommensurate structures at small twist angles). Finally, $\gamma_{\nu}(\omega,\mathbf{q})$ is a memory matrix that introduces the effect of harder modes of the stacking dynamics. The $\omega=0$, $\mathbf{q}=0$ limit of the memory matrix is a kinetic coefficient describing the friction between the two layers. It can be written as a Green-Kubo formula of the form\cite{phasons}\begin{align}
\gamma & \equiv \lim_{\omega\rightarrow 0}\lim_{\mathbf{q}\rightarrow 0}\gamma_{\nu}\left(\omega,\mathbf{q}\right)\\
& =\frac{1}{k_BT}\lim_{\epsilon\rightarrow 0^+}\lim_{\mathbf{q}\rightarrow 0}\int_0^{\infty}dt\,e^{-\epsilon t}\left\langle f_{\nu}\left(t,\mathbf{q}\right) f_{\nu}\left(0,\mathbf{q}\right) \right\rangle_T,
\nonumber
\end{align}
where I have introduced equilibrium correlation functions between microscopic forces $f_{\nu}(t,\mathbf{q})$ acting on mode $(\nu,\mathbf{q})$. The important observation is that the effect of these fluctuating forces is finite at $\mathbf{q}=0$ in general due to anharmonic coupling with optical phonons, reflecting that the linear momentum of an individual layer is not conserved; $\gamma$ is its characteristic relaxation rate. Consequently, under the application of a mechanical force the layer slides at some finite velocity in the steady state; otherwise, the layer would accelerate infinitely. From the spectral point of view, this means that phasons are overdamped at long wavelengths.\cite{phasons} %, and the stacking response to long-wavelength dynamical forces is diffusive rather than through propagating waves.\cite{phasons}

Consider first the pumping scenario, i.e., the flow of electric charge due to a mechanical force in the absence a voltage bias, $\boldsymbol{\mathcal{E}}(\omega)=0$. In this case the constitutive relations dictate\begin{subequations}\begin{align}
& \boldsymbol{j}\left(\omega\right)=\hat{A}\left(\omega\right)\cdot\boldsymbol{f}\left(\omega\right),\\
& \boldsymbol{v}\left(\omega\right)=-i\omega\,\hat{\chi}\left(\omega\right)\cdot\boldsymbol{f}\left(\omega\right)
\end{align}
\end{subequations}
Note that the $\omega=0$ limit in the second relation reads $\boldsymbol{v}(0)=\boldsymbol{f}(0)/\gamma\varrho$, i.e., one layer slides at the velocity imposed by the applied force compensated by the friction with the other layer. Combining both equations we have \begin{align}
\boldsymbol{j}\left(\omega\right)=\frac{i}{\omega}\hat{A}\left(\omega\right)\cdot\hat{\chi}^{-1}\left(\omega\right)\cdot\boldsymbol{v}\left(\omega\right).
\end{align}
Comparing Eq.~\eqref{eq:constitutive1} with this last equation we conclude that at small frequencies $\hat{A}(\omega)$ should go as \begin{align}
A_{ij}\left(\omega\right)=\frac{\varrho_{\phi}^{-1}}{\gamma-i\omega}\left(C_{\parallel}\delta_{ij}-C_{\perp}\epsilon_{ij}\right).
\end{align}

Now we consider the reciprocal effect, i.e., the sliding of one layer with respect to the other due to a voltage bias in the absence of a mechanical force, $\boldsymbol{f}(\omega)=0$. By making use of the last result along with Onsager reciprocity (Eq.~\ref{eq:Onsager}), the constitutive relations dictate\begin{align}
\boldsymbol{v}\left(\omega\right)=\frac{\varrho_{\phi}^{-1}}{\gamma-i\omega}\left(C_{\parallel}-C_{\perp}\hat{\mathbf{z}}\times\right)\boldsymbol{\mathcal{E}}\left(\omega\right).
\end{align}
The electric field makes the layers to slide as if they were subjected to the shear force $\boldsymbol{f}$ in Eq.~\eqref{eq:constitutive2}. Alternatively, in the presence of a bias voltage if the layers do not slide there must be a mechanical force opposing Eq.~\eqref{eq:constitutive2} so that the relative position of the layers remains locked. We will come back to this scenario later.

Finally, note that thermodynamic stability imposes the following relation between transport coefficients,\begin{align}
\label{eq:stability}
\frac{\gamma\varrho_{\phi}}{\rho}\geq C_{\parallel}^2+C_{\perp}^2,
\end{align}
where $\rho=\sigma^{-1}(0)$ is the electric resistivity.

\subsection{Experimental scenarios}

Let us consider first the scenario in which the device geometry (due to the contacts, encapsulation, etc.) prevents the two layers to move with respect to each other, $\boldsymbol{v}(\omega)=0$. %, but in principle charge can flow between contacts.
The constitutive relations dictate \begin{subequations}\begin{align}
& \boldsymbol{j}\left(\omega\right)=\hat{\sigma}\left(\omega\right)\cdot \boldsymbol{\mathcal{E}}\left(\omega\right)+\frac{C_{\parallel}+C_{\perp}\hat{\mathbf{z}}\times}{\gamma\varrho\left(1-i\omega\gamma^{-1}\right)}\,\boldsymbol{f}\left(\omega\right),\\
& \left(C_{\parallel}-C_{\perp}\hat{\mathbf{z}}\times\right)\boldsymbol{\mathcal{E}}\left(\omega\right)+\boldsymbol{f}\left(\omega\right)=0.
\end{align}
\end{subequations}
By bringing the the value of the force deduced from the second relation to the first equation we have\begin{subequations}\begin{align}
& \boldsymbol{j}\left(\omega\right)=\left[\hat{\sigma}\left(\omega\right)-\frac{\sigma_0}{1-i\omega\gamma^{-1}}\right] \boldsymbol{\mathcal{E}}\left(\omega\right),\\
&\textrm{with}\,\,\,\sigma_0=\frac{C_{\parallel}^2+C_{\perp}^2}{\gamma\varrho_{\phi}}.
\end{align}
\end{subequations}
We find a Drude-like negative contribution to the effective conductivity. When the Fermi level lies within the electron gap we expect $\hat{\sigma}(0)\rightarrow 0$ ($\rho\rightarrow \infty$) at low temperatures and it seems that the effective conductivity is dominated by this new negative correction given by\begin{align}
\sigma_0=\frac{ne^2\tau}{m_{*}},\,\,\, \textrm{with}\,\,\, m_*=\frac{\theta^2\varrho_{\phi}}{n},
\end{align}
where $\tau\equiv\gamma^{-1}$ is the momentum relaxation time. Note here the similarity of this expression with the sliding conductivity of an incommensurate charge density wave,\cite{RiceAndersonLee} where $m_{*}$ defined above plays the role of the effective mass of the sliding charge.

The important caveat is that nothing is really sliding here and this apparent charge counterflow should be absent. The resolution of this paradox is that phasons are no longer soft, there must be a pinning gap $\omega_{\textrm{p}}$ in the spectrum, and we should replace the Drude-like factor $(\gamma-i\omega)^{-1}$ by $-i\omega/(\omega_{\textrm{p}}^2-\omega^2-i\omega\gamma)$ in these expressions. There is no DC transport in linear response, but there is a correction to the electric polarization density $\mathbf{P}(\omega)=-\boldsymbol{j}(\omega)/i\omega z_0$ (defined here per volume, with $z_0$ being the thickness of the bilayer) opposing the applied field, or in other words, a negative correction to the effective electric susceptibility,\begin{subequations} \begin{align}
\delta\chi_e\left(\omega\right) = & -\frac{C_{\parallel}^2+C_{\perp}^2}{\epsilon_0z_0\varrho_{\phi}\left(\omega_{\textrm{p}}^2-\omega^2-i\omega\gamma\right)}\\&\xrightarrow[\omega\rightarrow 0]{}-\frac{n e^2}{\epsilon_0z_0m_*\omega_{\textrm{p}}^2},
\end{align} \end{subequations}
%The last result holds for twisted bilayer graphene in the adiabatic limit.
where $\epsilon_0$ is vacuum permittivity. The pinning gap is ultimately related to a smooth (so that the gap remains open) distortion of the moir\'e pattern under the application of an electric field. In this case the adiabatic approximation is ensured so long $\omega_{\textrm{p}}^{-1}$ is long compared to the time scale set by the electronic gap.

%Elaborate on instabilities? Elaborate on minimal value for friction

We can consider the alternative situation of a purely mechanical experiment/device in which the layers are free to slide ($\omega_{\textrm{p}}=0$) but there are no electric contacts, so charge cannot flow, $\boldsymbol{j}=0$. In this case the constitutive relations in the $\omega=0$ limit %read\begin{subequations}\begin{align}
%& 0=\boldsymbol{\mathcal{E}}+\frac{\rho}{\gamma\varrho}\left(C_{\parallel}+C_{\perp}\hat{\mathbf{z}}\times\right)\boldsymbol{f},\\
%& \gamma\varrho \boldsymbol{v}=\boldsymbol{f}_{\boldsymbol{\mathcal{E}}}+\boldsymbol{f}.
%\end{align}
%\end{subequations}
%These relations include the mechanical force created by the electric field $\boldsymbol{\mathcal{E}}$ built from charge accumulation in the steady state, which is in turn proportional to the applied force. Combining both equations we have 
dictate that layers slide at a rate $\boldsymbol{v}=\boldsymbol{f}/\tilde{\gamma}\varrho$ with an enhanced friction parameter $\tilde{\gamma}$ given by\begin{align}
\tilde{\gamma}=\frac{\gamma}{1-\sigma_0\rho},
\end{align}
where $\sigma_0$ is the \textit{sliding conductivity} introduced before. Note that $1\geq \sigma_0\rho \geq0$, hence $\tilde{\gamma}>0$ for thermodynamic stability, Eq.~\eqref{eq:stability}.

This last result expresses that the friction between layers is effectively larger when the system is charged. As the layers slide there is a pumping electromotive force perpendicular (in the case of twisted bilayer graphene) to the motion. If the system is isolated (disconnected from reservoirs/contacts) the pumped charge cannot leave the system, so a charge accumulation builds up. The associated voltage drop cancels the pumping electromotive force in the steady state. But this voltage drop produces an additional mechanical force opposing the original driving force, hence modifying the relation between force and sliding velocity: one has to apply a larger force to slide at the same velocity. Quantum mechanically, the spectral flow linked to this charge accumulation implies the existence of edge states associated with the moir\'e pattern.\cite{edge1,edge2,edge3} %Note here the analogy with Hall physics, where changes in the stacking configuration are akin to the adiabatic insertion of magnetic flux.

%Elaborate on charge accumulation, edge states, manifestation of the counter force produce by the voltage drop.

\section{Depinning fields}

\label{sec:depinning}

The main result in Eq.~\eqref{eq:constitutive2} suggests that if external conditions do not prevent the sliding motion of the layers (the second scenario discussed in the previous section) an arbitrary small field should be able to drive this motion. In real devices, however, the presence of disorder implies a finite threshold that the driving force must overcome first. A model for the locked-to-sliding transition is beyond the linear response theory presented here. The following is a rough estimate of the order of magnitude of the depinning fields in the same spirit as in the problem for charge density waves.\cite{Lee1,Lee2}

In order to estimate the total pinning force opposing the driving field I am going to neglect thermal fluctuations and consider only the static mechanical response to a quenched distribution of random forces acting on stacking configurations. The mechanical free energy is $\mathcal{U}+\int d\mathbf{r}\, \mathcal{V}_{\textrm{dis}}(\mathbf{r},\delta\boldsymbol{\phi})$; the disorder potential $\mathcal{V}_{\textrm{dis}}(\mathbf{r},\delta\boldsymbol{\phi})$ favors deviations of the stacking order at a given point $\mathbf{r}$ from the one imposed by the moir\'e pattern and the mutual interaction between layers, therefore pinning the relaxed texture $\boldsymbol{\phi}_0(\mathbf{r})$. We assume that disorder is weak, in the sense that spatial variations of stacking fields are smooth compared to the moir\'e pitch. We can then expand to linear order in $\delta\boldsymbol{\phi}$, $\mathcal{V}_{\textrm{dis}}(\mathbf{r},\delta\boldsymbol{\phi})\approx-\boldsymbol{f}_{\textrm{dis}}(\mathbf{r})\cdot\delta\boldsymbol{\phi}$, where $\boldsymbol{f}_{\textrm{dis}}(\mathbf{r})$ is some distribution of microscopic pinning forces. For simplicity, we model them with a Gaussian distribution with variance\begin{align}
\left\langle f_i(\mathbf{r}) f_j(\mathbf{r}') \right \rangle_{\textrm{dis}}=\overline{f^2_{\textrm{dis}}}\,\xi^2\,\delta_{ij}\, \delta\left(\mathbf{r}-\mathbf{r}'\right).
\end{align}
The model can be understood as spatial distribution of pinning centers of characteristic force $(\overline{f^2_{\textrm{dis}}})^{1/2}$ coarse-grained in a microscopic length $\xi$ representing its characteristic range.

%The problem consists of summing up these microscopic forces. %The important observation is that 
From general arguments for elastic media,\cite{Imry-Ma,Larkin} the moir\'e pattern always breaks into finite domains regardless of the microscopic origin of disorder.\cite{phasons} %I will refer to the characteristic size of these domains as the collective pinning length $L_{\textrm{pin}}$.\cite{Lee1,Larkin} 
Beyond its characteristic size $L_{\textrm{pin}}$ the stacking configurations are no longer correlated and the cumulative effect of the microscopic pinning forces must stop. %We can estimate the strength of the collective force as\begin{align}
%\boldsymbol{F}_{\textrm{pin}}^2=\int_{L_{\textrm{pin}}^2}\int_{L_{\textrm{pin}}^2}\left\langle \boldsymbol{f}(\mathbf{r})\cdot \boldsymbol{f}(\mathbf{r}') \right \rangle_{\textrm{dis}}=\overline{f^2_{\textrm{dis}}}\xi^2L_{\textrm{pin}}^2.
%\end{align}
Due to their random orientation the total pining force on the domain grows with the length, not the area. This opposes the driving force, Eq.~\eqref{eq:constitutive2} integrated over the area $L_{\textrm{pin}}^2$. Consequently, there is always a finite strength of the electric field that overcomes pinning, \begin{align}
\mathcal{E}_c=\frac{\xi}{L_{\textrm{pin}}}\sqrt{\frac{\overline{f^2_{\textrm{dis}}}}{C_{\parallel}^2+C_{\perp}^2}}.
\end{align}%The remaining task is to compute the collective pinning length $L_{\textrm{pin}}$. 
The size of the domains corresponds to the distance at which the spatial fluctuations in stackings exceed the microscopic length $\xi$. %, $\langle (\delta\boldsymbol{\phi}(L_{\textrm{pin}})-\delta\boldsymbol{\phi}(0))^2\rangle_{\textrm{dis}}\sim \xi$.
In linear response, taking the static limit $\omega=0$ of the susceptibility in Eq.~\eqref{eq:susceptibility} and dropping the logarithm in the estimate of $L_{\textrm{pin}}$,\cite{phasons} we have %the correlator can be estimated as\begin{align}
%\nonumber
%\left\langle \left(\delta\boldsymbol{\phi}(L_{\textrm{pin}})-\delta\boldsymbol{\phi}(0)\right)^2\right\rangle_{\textrm{dis}}\approx & \frac{8L_{\textrm{pin}}^2\overline{f^2_{\textrm{dis}}}\xi^2}{\kappa_L^2\kappa_T^2}\left(\kappa_L^2+\kappa_T^2\right)\\
%& \times\int_{\frac{L_{\textrm{pin}}}{L}}^{\infty}d\zeta\,\frac{1-J_0\left(\zeta\right)}{\zeta^3},
%\end{align}
%, $J_0(\zeta)$ is a Bessel function, and the logarithmic divergence of the last integral has been cut-offed by the size of the sample $L$. The rapid growth of the disorder correlators gives rise to the lack of positional order in the moir\'e superlattice. Dropping the logarithm for simplicity we have \begin{align}
%L_{\textrm{pin}}\approx\sqrt{\frac{\pi}{\overline{f^2_{\textrm{dis}}}}}\frac{\kappa_L\kappa_T}{\sqrt{\kappa_L^2+\kappa_T^2}},
%\end{align}
%and therefore
\begin{align}
\mathcal{E}_c\approx\sqrt{\frac{3\left(\kappa_L^2+\kappa_T^2\right)}{\pi}}\,\frac{a^2\overline{f^2_{\textrm{dis}}}\xi}{8e\theta\kappa_L\kappa_T},
\end{align}
where $\kappa_{L,T}$ is the stiffness of longitudinal and transverse phason fluctuations and I have used the result for twisted bilayer graphene in Eq.~\eqref{eq:estimate}.

Taking $\xi\sim a$ and disorder strength of the order of adhesion energies between graphene layers, $\xi(\overline{f^2_{\textrm{dis}}})^{1/2}\sim 4$ meV/$\AA^2$ (the difference between AA and AB/BA stacking configurations\cite{VdW_energies}), and identifying $\kappa_L=\mu$, $\kappa_T=\lambda+2\mu$, where $\lambda=3.25$ eV/$\AA^2$, $\mu=9.57$ eV/$\AA^2$ are graphene's Lamé coefficients,\cite{Lame} the collective pinning lengths are estimated in $L_{\textrm{pin}}\sim500$ nm, which is roughly the characteristic size of domains in the landscape of twist angle variations around the magic condition observed in experiment.\cite{tomography} Bringing these numbers to the expression of the depinning field we have $\mathcal{E}_c\sim 5$ kV/cm at the magic angle. Note, however, that for the same disorder parameters the depinning fields are smaller for larger twist angles, $\mathcal{E}_c\propto \theta^{-1}$.

Are these small or large? We can compare them with the depinning fields for charge density waves. Naturally these vary from one material to another depending on the amount of disorder, temperature and other factors. In a typical system such as NbSe$_3$ threshold fields for sliding conduction range from tens of mV/cm\cite{NbSe3-1} to few V/cm.\cite{NbSe3-2} Threshold fields of few V/cm have been also reported recently for the axion insulator candidate (TaSe$_4$)$_2$I\cite{axion} and previously for a commensurate charge density wave in a manganite.\cite{manganite} We can conclude that the estimated $\mathcal{E}_c$ here is at least 2-3 orders of magnitude larger than the typical threshold fields observed in sliding conduction,\cite{RMP} with the recent exception of a nearly-commensurate 2D charge density wave in 1T-TaS$_2$, which is as large as 1 kV/cm.\cite{TaS2}

\section{Conclusions}

\label{sec:conclusions}

The constitutive relations written above describe the coupled response of a twisted Van der Waals bilayer against electric fields and layer-shear mechanical forces. These relations are general in the sense that they account also for intrinsic forces between layers responsible for lattice relaxation, pinning and friction. Along with the resistivity $\rho$ and the friction parameter $\gamma$ we must consider also new coefficients $C_{\parallel,\perp}$ describing reactive forces on stacking configurations exerted by an electric field when the moir\'e pattern is charged. From Onsager reciprocity these must be the same coefficients describing charge pumping by the sliding motion of one layer with respect to the other.\cite{pumping1,pumping2,pumping3}

This new mechanical force has a geometrical origin. In agreement with Refs.~\onlinecite{pumping1,pumping2,pumping3} the coefficients $C_{\perp,\parallel}$ appearing in the lowest-order correction to the mechanical action can be expressed as integrals of a Berry curvature defined on the mixed space spanned by momenta and stacking configurations. In twisted bilayer graphene $C_{\parallel}=0$ due to C$_2$ symmetry and $C_{\perp}=-e\nu/\theta A_{\textrm{m}}$, where $\nu$ is the band filling measured from neutrality. This is strictly true for completely filled/empty bands, which is a consequence of a topological quantization in terms of 4 sliding Chern numbers defined on this mixed space.\cite{pumping1,pumping2,pumping3} The calculation for noninteger fillings including a self-consistent Hartree potential shows that $C_{\perp}$ tracks the electron density for most fillings, although it is important to note that in the presence of a Fermi surface the adiabatic regime is not rigorously defined and there must be corrections to the force in Eq.~\eqref{eq:constitutive2}, including dissipative terms due to electron quasiparticle excitations. These processes introduce additional relaxation channels for the linear momentum of individual layers, and hence electronic contributions to interlayer friction.

The application of the constitutive relations shows that when the relative position of the layers is pinned there is still a negative correction to the electric polarization opposing the electric field. The reciprocal effect is an effective increase in the friction between layers when the system is charged and free to slide. This, however, might be challenging to probe in the lab due to the thermodynamic instability intrinsic to incommensurate layers.

%It is important to stress that a finite sliding conductivity when $\varrho\rightarrow\infty$ is incompatible with thermodynamic stability, Eq.~\eqref{eq:stability}. Moreover, this relation imposes a stringent condition for the observation of a topological Thouless pump, in the sense that when the chemical potential lies within the gap of the electron bands and $\varrho\rightarrow\infty$ the layers cannot be free to slide, at least not smoothly. Otherwise, the coupled charge-stacking dynamics would be energetically unstable. In this regard, charged impurities may contribute to pin the moir\'e pattern in the same way as in a Wigner crystal \cite{Wigner}.

It is important to stress that already in linear response thermodynamic stability (Eq.~\ref{eq:stability}) imposes stringent condition for the observation of a topological Thouless pump, in the sense that when the chemical potential lies within the gap of the electron bands and $\varrho\rightarrow\infty$ at low temperatures the layers cannot be free to slide. In this regard, charged impurities may contribute to pin the moir\'e pattern in the same way as in Wigner crystals \cite{Wigner} provided that charge is more concentrated in some stacking areas within the moir\'e cell.

Finally, for a model of weak disorder compatible with the observed deviations in twist angles close to the magic-angle condition, the depinning fields for the sliding motion of the graphene layers are of the order of $\mathcal{E}_c\approx 5$ kV/cm. These are large if we compare them with typical depinning fields for incommensurate charge density waves.\cite{RMP} This probably constitutes the major limitation for the control of the stacking order by electric means.%, which adds another challenge to the ability to control the stacking order by electric means in moir\'e systems.

%FERROELECTRIC INSTABILITIES?

%Is incommensurability required for adiabatic pumping? I don't think so, it is just enough to change adiabatically the stacking configuration, that does not requiere the system to explore all stacking configurations at the same time (in that case the pumping current would not be constant in time). In fact, that is Koshino paper. The problem is: how do you realize this in practice? We need the moir\'e pattern to \textit{flow} (the commensurate cell is some integer multiple of the moire pattern), but it flows only if it is unpinned. 

%Adiabatic pumping requires that changes in stacking are cyclic (as well as a band structure), no incommensurability, and bands (or a notion of BZ). Does a constitutive relation require commensurability? A moir\'e flow requieres commensurability.

\begin{acknowledgments}
I am grateful to R. M. Fernandes and to D. Xiao for useful correspondence, and to F. de Juan for pointing me to relevant literature on charge density waves. Funding from the Spanish MCI/AEI/FEDER through Grant No. PID2021-128760NB-I00 is acknowledged.
\end{acknowledgments}

\appendix

\section{Sliding Berry curvature from Eq.~\eqref{eq:index}}

\label{sec:app1}

The free Green operator in the Bloch basis $|n,\mathbf{q},\alpha\rangle$, where $n$ is the band index and $\alpha$ represents the rest of quantum numbers (spin, valley, etc.), can be written as\begin{align}
\hat{\mathcal{G}}_0\left(\omega,\mathbf{q}\right)= & \sum_{n,\alpha}\left[\frac{\Theta \left(\varepsilon_{n,\alpha}(\mathbf{q})-\mu\right)}{\hbar\omega+i 0^+-\left(\varepsilon_{n,\alpha}(\mathbf{q})-\mu\right)}+\right.\\
& \left.
\frac{\Theta \left(\mu-\varepsilon_{n,\alpha}(\mathbf{q})\right)}{\hbar\omega+i 0^--\left(\varepsilon_{n,\alpha}(\mathbf{q})-\mu\right)}\right]\left|n,\mathbf{q},\alpha \right\rangle  \left\langle n,\mathbf{q},\alpha \right|.
\nonumber
\end{align}
$\varepsilon_{n,\alpha}(\mathbf{q})$ is the band dispersion and $\Theta(x)$ is the Heaviside step function. Hereafter the electronic Hamiltonian is assumed to be diagonal in $\alpha$ numbers. Equation~\eqref{eq:index} can be written as\begin{widetext}\begin{align}
\nonumber
\tilde{C}_{ij}= & \sum_{\alpha}\sum_{n_1,n_2}\int_{\textrm{mBZ}}\frac{d\mathbf{q}}{(2\pi)^2}\,\left\langle n_1,\mathbf{q},\alpha\right| \partial_{\phi_j}\hat{\mathcal{H}}_0\left|n_2,\mathbf{q},\alpha \right\rangle \left\langle n_2,\mathbf{q},\alpha\right| \partial_{q_i}\hat{\mathcal{H}}_0 \left| n_1,\mathbf{q},\alpha \right\rangle \int_{\infty}^{\infty}\frac{d\hbar\omega}{2\pi}\, \left(\frac{\Theta \left(\varepsilon_{n_1,\alpha}(\mathbf{q})-\mu\right)}{\hbar\omega+i 0^+-\left(\varepsilon_{n_1,\alpha}(\mathbf{q})-\mu\right)}
\right.\\
& \left.
+ \frac{\Theta \left(\mu-\varepsilon_{n_1,\alpha}(\mathbf{q})\right)}{\hbar\omega+i 0^--\left(\varepsilon_{n_1,\alpha}(\mathbf{q})-\mu\right)}\right)^2\left(\frac{\Theta \left(\varepsilon_{n_2,\alpha}(\mathbf{q})-\mu\right)}{\hbar\omega+i 0^+-\left(\varepsilon_{n_2,\alpha}(\mathbf{q})-\mu\right)}
+ \frac{\Theta \left(\mu-\varepsilon_{n_2,\alpha}(\mathbf{q})\right)}{\hbar\omega+i 0^--\left(\varepsilon_{n_2,\alpha}(\mathbf{q})-\mu\right)}\right).
\end{align}
Only when the two band poles lie at different halves of the complex plane the result of the integral in $\omega$ is different form $0$. The result is\begin{align}
\nonumber
\tilde{C}_{ij} & =\sum_{\alpha}\sum_{n_1,n_2}\int_{\textrm{mBZ}}\frac{d\mathbf{q}}{(2\pi)^2}\,\frac{-2\,\textrm{Im}\left[\left\langle n_1,\mathbf{q},\alpha\right| \partial_{\phi_j}\hat{\mathcal{H}}_0\left|n_2,\mathbf{q},\alpha \right\rangle \left\langle n_2,\mathbf{q},\alpha\right| \partial_{q_i}\hat{\mathcal{H}}_0 \left| n_1,\mathbf{q},\alpha \right\rangle\right]}{\left(\varepsilon_{n_1,\alpha}(\mathbf{q})-\varepsilon_{n_2,\alpha}(\mathbf{q})\right)^2}\,\Theta\left(\mu-\varepsilon_{n_2,\alpha}(\mathbf{q})\right)\Theta\left(\varepsilon_{n_1,\alpha}(\mathbf{q})-\mu\right)\\
& = \sum_{\alpha,n}\int_{\textrm{mBZ}}\frac{d\mathbf{q}}{(2\pi)^2}\,\Omega_{q_i\phi_j}^{(n,\alpha)}\left(\mathbf{q}\right)\,\Theta\left(\mu-\varepsilon_{n,\alpha}(\mathbf{q})\right),
\end{align}
where I have introduced the sliding Berry curvature\cite{pumping1,pumping2,pumping3}\begin{align}
\Omega_{q_i\phi_j}^{(n,\alpha)}\left(\mathbf{q}\right)=\sum_{n'\neq n}\frac{-2\,\textrm{Im}\left[\left\langle n,\mathbf{q},\alpha\right| \partial_{q_i}\hat{\mathcal{H}}_0\left|n',\mathbf{q},\alpha \right\rangle \left\langle n',\mathbf{q},\alpha\right| \partial_{\phi_j}\hat{\mathcal{H}}_0 \left| n,\mathbf{q},\alpha \right\rangle\right]}{\left(\varepsilon_{n,\alpha}(\mathbf{q})-\varepsilon_{n',\alpha}(\mathbf{q})\right)^2}.
\end{align}
\end{widetext}

\section{Self-consistent Hartree potential}

\label{sec:app2}

The Hartree potential (diagonal in sublattices and layers) reads\begin{align}
V_{\textrm{H}}\left(\mathbf{r}\right)=\int d\mathbf{r}'\, \frac{e^2}{4\pi\epsilon_0\epsilon_r\left|\mathbf{r}-\mathbf{r}'\right|}\, n\left(\mathbf{r}'\right),
\end{align}
where $\epsilon_r$ describes the dielectric environment. The electron density is periodic on the moir\'e pattern and thus admits a Fourier expansion of the form\begin{align}
n\left(\mathbf{r}\right)=\sum_{\left\{\mathbf{G}\right\}}n_{\mathbf{G}}\, e^{i\mathbf{G}\cdot\mathbf{r}},
\end{align}
where $n_{\mathbf{G}}$ are computed self-consistently from the Hartree bands $\varepsilon_{n,\zeta}^{\textrm{H}}(\mathbf{q})$ and wave functions $u_{n,\mathbf{q},\zeta}^{\textrm{H}}(\mathbf{r})$, \begin{align}
n_{\mathbf{G}}=4\sum_{n}\int_{\textrm{mBZ}}\frac{d\mathbf{q}}{(2\pi)^2}\,\mathcal{f}_{\mathbf{G}}^{(n,+)}\left(\mathbf{q}\right)\Theta\left(\mu-\varepsilon_{n,\zeta}^{\textrm{H}}(\mathbf{q})\right).
\end{align}
The factor 4 comes from spin and valley degeneracies, and I have introduced the form factors\begin{align}
\mathcal{f}_{\mathbf{G}}^{(n,\zeta)}\left(\mathbf{q}\right)=\int d\mathbf{r}\, e^{-i\mathbf{G}\cdot\mathbf{r}}\left[u_{n,\zeta,\mathbf{q}}^{\textrm{H}}\left(\mathbf{r}\right)\right]^*u_{n,\zeta,\mathbf{q}}^{\textrm{H}}\left(\mathbf{r}\right).
\end{align}

In the calculations of Fig.~\ref{fig:fig3} the number of harmonics in the Fourier expansion of $n(\mathbf{r})$ was restricted to momenta $\mathbf{G}$ in the first star of the moir\'e; $n_{\mathbf{G}}$ reduces then to the same real number for the six vectors due to $D_6$ symmetry. The bands in Fig.~\ref{fig:fig3} were computed after determining this number in an iterative calculation with an error inferior to 1\% for these two examples (errors never exceeded 2.5\% in the calculations of Fig.~\ref{fig:fig4}). The appreciable distortion of the flat band dispersion agrees well with similar calculations in the literature.\cite{Hartree1,Hartree2,Hartree3,Hartree4}

For the calculation of the matrix elements of the electron-phason coupling is important to emphasize again that all position-dependent terms in the Hamiltonian are a functional of the stacking configuration and, in particular, there is also a contribution from the Hartree term. These expressions for the Hartree potential were implicitly written in the coordinate frame defined by $\boldsymbol{\phi}_0(\mathbf{r})$. Under an adiabatic change in stacking configurations the wave functions and therefore the form factors are transformed accordingly, $\mathcal{f}_{\mathbf{G}}^{(n,\zeta)}(\mathbf{q})\rightarrow e^{i\theta\left(\mathbf{G}\times\delta\boldsymbol{\phi}\right)_z}\mathcal{f}_{\mathbf{G}}^{(n,\zeta)}(\mathbf{q})$. Thus, the Hartree potential can be written as a functional of the stacking field as\begin{align}
V_{\textrm{H}}\left(\mathbf{r}\right)\equiv V_{\textrm{H}}\left[\boldsymbol{\phi}\left(\mathbf{r}\right)\right]=\sum_i e^{i\mathbf{g}_i\cdot\boldsymbol{\phi}(\mathbf{r})}\frac{n_{\mathbf{G}_i}e^2}{2\epsilon_0\epsilon_r\left|\mathbf{G}_i\right|}.
\end{align}


\begin{thebibliography}{99}
\bibitem{foot_rot} The convention is that the top layer is rotated anti-clockwise with respect to the bottom layer an angle $\theta$.
\end{thebibliography}
\end{document}